\documentclass{article}

\usepackage{PRIMEarxiv}

\usepackage[utf8]{inputenc} 
\usepackage[T1]{fontenc}    
\usepackage{hyperref}       
\usepackage{url}            
\usepackage{booktabs}       
\usepackage{amsmath}
\usepackage{subcaption}
\usepackage{amsfonts}       
\usepackage{nicefrac}       
\usepackage{microtype}      
\usepackage{lipsum}
\usepackage{fancyhdr}       
\usepackage{graphicx}       
\graphicspath{{media/}}     

\title{Temperature-Dependent Evolution of Coherence, Entropy, and Photon Statistics in Photoluminescence}

\author{
  Tomer Bar Lev \\
 Faculty of Mechanical Engineering, Technion—Israel Institute of Technology \\
  Haifa 32000, Israel\\
   \And
  Carmel Rotschild \\
   Faculty of Mechanical Engineering, Technion—Israel Institute of Technology \\
   Russell Berrie Nanotechnology Institute, Technion—Israel Institute of Technology \\
    Haifa 32000, Israel\\\\
  \texttt{carmelr@technion.ac.il} \\
}

\begin{document}
\maketitle

\begin{abstract}
Photoluminescence (PL) is a fundamental light–matter interaction in which absorbed photons are re-emitted, playing a key role in science and engineering. It is commonly modeled by introducing a non-zero chemical potential into Planck’s law to capture its deviation from thermal emission. In this work, we establish, for the first time to our knowledge, a fundamental relationship that expresses the chemical potential as a function of temperature, material properties, and excitation conditions, enabling a treatment of PL analogous to Planck’s law with thermal radiation. This formulation allows for the analysis of temperature-dependent PL properties, including spectral emission, entropy, temporal coherence, and photon statistics, capturing the transition from narrowband pump-induced to broadband thermal emission. 
Notably, we identify a temperature range where the emission rate is quasi-conserved, associated with the previously reported blueshift. This is followed by a rapid transition to thermal behavior, reflected in both the chemical potential and entropy. Conversely, the coherence time and photon statistics evolve smoothly across the entire temperature range. 
Alongside its scientific contribution, this framework provides a foundation for designing temperature-tunable light sources, enabling control over coherence length and photon statistics.
\end{abstract}

\section{Introduction}
Central to PL is the principle of energy conservation. When a high-energy photon is absorbed, an electron is excited across the bandgap following a fast thermalization, and ultimately, the electron recombines, typically emitting a red-shifted photon. This difference in absorbed and emitted spectra is referred to as the Stokes shift \cite{lakowicz2006principles,valeur2002molecular}.

Another intriguing aspect of PL emission is photon-rate conservation\cite{manor2015conservation,Kurtulik2025}, which is defined by the quantum efficiency $(QE)$, a material property describing the probability of emitting an absorbed photon at low temperatures. PL occurs through external excitation and thermal radiation, governed by the material’s phonon temperature. Any form of radiation, and PL in particular, can be described by the generalized Planck’s law, providing temperature-dependent thermal contribution enhanced by the chemical potential, $\mu$ \cite{wurfel1982chemical, feuerbacher1990verification, ries1991chemical, ross1967some, yablonovitch1995light, manor2015conservation, Kurtulik2025, wurfel1995generalized}:

\begin{equation} \label{eq. Generalized Planck}
R \left( \nu,T,\mu \right) = \alpha(\nu) \cdot \frac{2\nu^2}{c^2} \frac{1}{e^{\frac{h\nu-\mu}{kT}}-1} \approx \alpha(\nu) \cdot  R_{BB}(\nu,T) \cdot e^{\frac{\mu}{kT}}
\end{equation}

where $R$ is an arbitrary spectral photon flux (photons per second, per frequency, per solid angle, per unit area), $R_{BB}$ is the spectral photon flux of a blackbody, $h\nu$ is the photon energy, $T$ is the material’s temperature, $\mu$ is the chemical potential, describing the excitation level above thermal emission, $\alpha$ is the absorptivity, and $kT$ is the thermal energy. The scalar chemical potential within a spectral band describes a PL spectrum that is similar to thermal emission scaled by $\mu$. 

Previous theoretical and experimental studies on temperature-dependent PL have shown that the photon emission rate remains quasi-conserved at low temperatures~\cite{manor2015conservation, Kurtulik2025}. As the temperature increases, this rate rises sharply, reaching a universal point where the PL photon rate equals the absorbed pump rate, as the PL body’s temperature matches the pump’s brightness temperature.  

While previous studies have focused on the temperature dependence of PL in specific materials and conditions~\cite{lu2014temperature, du2019temperature, Karmegam2023, Peng2022, zheng2014temperature}, this work aims to develop a generalized framework for PL across systems. To achieve this, we derive the inherent dependence of the chemical potential on the system parameters, allowing PL to be described with the same formalism as thermal emission. This finding will enable us to analyze further key properties of PL, such as the emission spectrum, entropy generation, temporal coherence, and photon statistics. In addition, we discuss the conditions for thermodynamic equilibrium between different pumps and the PL body. 

\section{Theoretical Framework}
By solving general rate equations at a specific frequency for a 2-level system, Kurtulik \textit{et al.} showed that PL emission rate evolves with temperature and is formulated by thermal and pump-induced terms\cite{Kurtulik2025}:
\begin{equation}
    \label{eq Kurtulik: R_th R_pump}
    \begin{array}
    {l}{R_{\text{thermal}}}\left( T \right) = \alpha  \times \left( {1 - QE} \right) \times {R_{BB}}\left( T \right)\\{R_{\text{pump-induced}}} = \alpha  \times QE \times {R_{pump}}\left( {{T_p}} \right)
    \end{array}
\end{equation}
Extending this solution to a continuous spectrum above the energy bandgap results in (see detailed derivation in Supplement 1): 
\begin{equation} \label{eq PL_broadband}
\begin{aligned}
    R_{PL}\left( {\nu ,T,T_p} \right) = \alpha \left( {1 - QE} \right) \cdot {R_{BB}}\left( {\nu ,T} \right) +
    \alpha  \cdot QE \cdot \int \limits_{(\nu)}  {{R_{pump}}\left( {\nu ,{T_p}} \right)d\nu }  \cdot \frac{{{R_{BB}}\left( {\nu ,T} \right)}}{{\int \limits_{(\nu)}  {{R_{BB}}\left( {\nu ,T} \right)d\nu } }}
    \end{aligned}
\end{equation}
where $(\nu)$ denotes the frequency interval beyond the bandgap $[\nu_{bg}, \infty)$, and ${\int \limits_{(\nu)}  {{R_{BB}}}} \cdot d\nu$, ${\int \limits_{(\nu)}  {{R_{pump}}}} \cdot d\nu$ are the total fluxes of the blackbody and the pump, respectively (photons per second, per solid angle, per unit area). The pump at the discussed spectral band is chosen to be a blackbody at $T_p$, which can be achieved by placing a band-pass filter between the thermal pump and the PL material. Like the 2-level system in Eq.~(\ref{eq Kurtulik: R_th R_pump}), the first term represents the thermal contribution of the emission, whereas the second accounts for externally excited (pumped) electrons. As shown by Kurtulik \textit{et al.}, in this case, there is a universal temperature, $T=T_p$, where the PL flux matches that of the absorbed pump \cite{Kurtulik2025}. Thus, the material behaves as a thermal body in thermal equilibrium with the pump, and Eq.~(\ref{eq PL_broadband}) is reduced to Planck’s law: $R_{PL}(\nu,T) = \alpha \cdot R_{BB}(\nu,T)$, independent of the $QE$. 

\begin{figure}[h!]
    \centering
    \begin{subfigure}[]{0.49\textwidth}
        \centering
        \includegraphics[width=\linewidth]{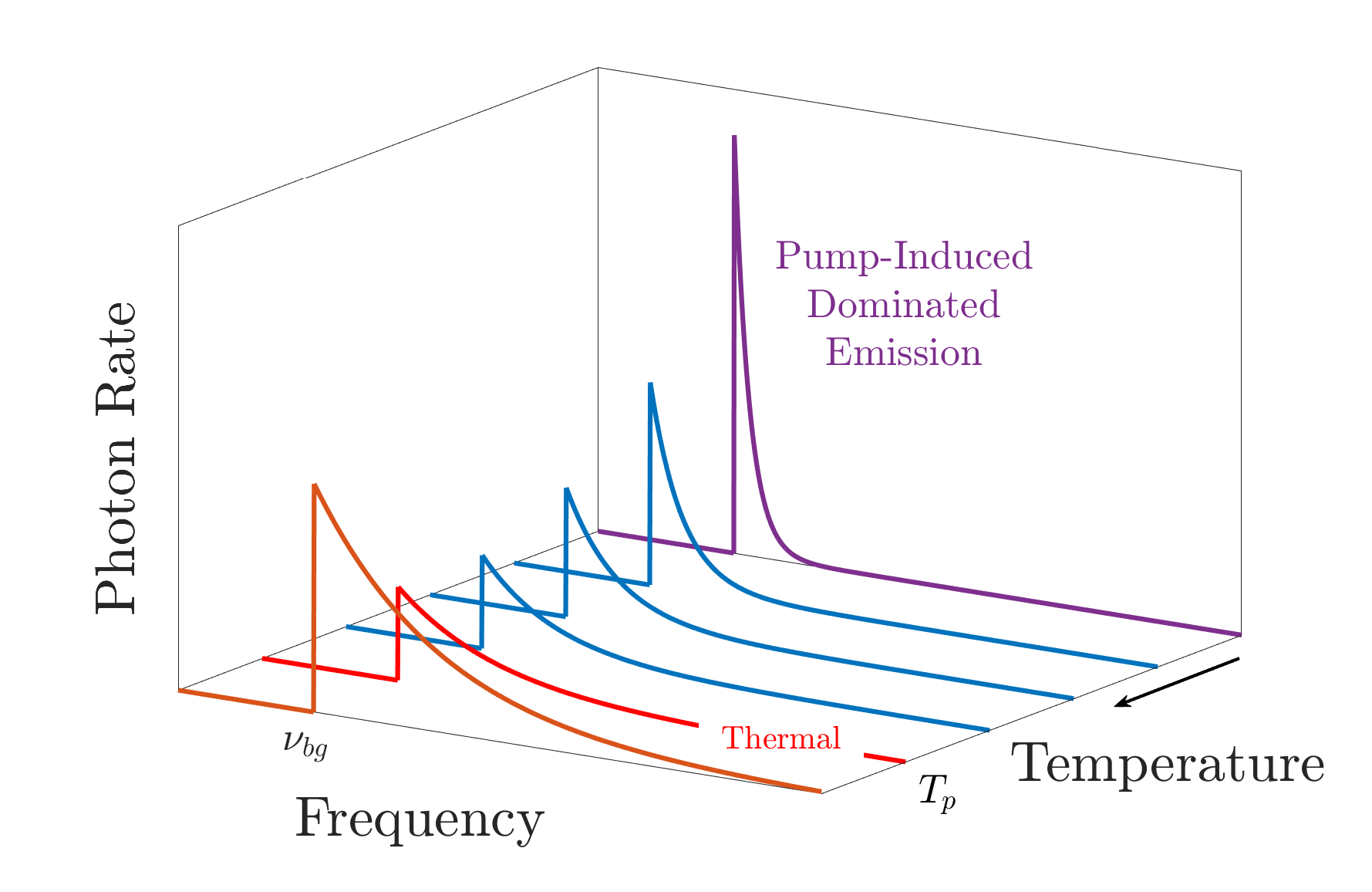}
        \caption{}
        \label{Fig1 spectra vs T}
    \end{subfigure}
    \hfill
    \begin{subfigure}[]{0.49\textwidth}
        \centering
        \includegraphics[width=\linewidth]{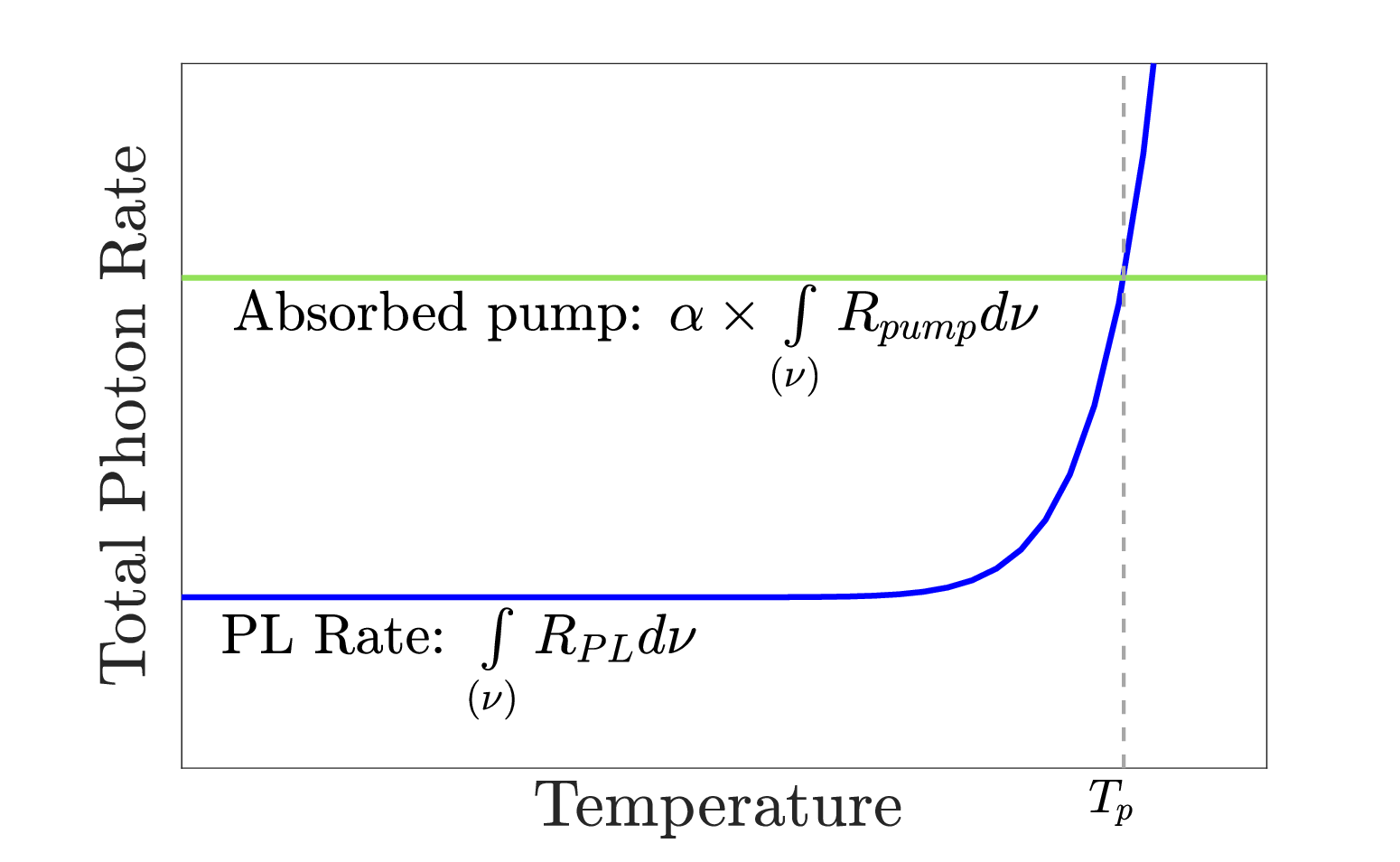}
        \caption{}
        \label{Fig1 rate vs T}
    \end{subfigure}
    \caption{Illustration of the photon spectra and total photon rate as a function of temperature. (a) Evolution of the spectral photon rate with temperature of a PL material pumped by a blackbody at $T_p$. At $T \ll T_p$, the emission is pump-dominated and spectrally narrow.  As $T$ increases, the rate at higher frequencies increases at the expense of the lower (blueshift), maintaining a quasi-conserved rate. Towards the universal temperature $(T=T_p)$, the emission becomes thermally dominated, increasing across all frequencies. (b) Total photon rate versus temperature (blue). At the universal temperature, the total photon rate equals the total absorbed pump rate (green).}
\end{figure}

The fast thermalization of excited electrons results in equilibrium with the lattice phonons, leading to the Boltzmann distribution~\cite{Siegman1986,andersson2020enhancing}. Therefore, their distribution, followed by the emission, are temperature-dependent, which can be seen in Eq.~(\ref{eq. Generalized Planck}).
As the temperature rises, the electron distribution shifts accordingly, while the thermal excitation remains negligible at low temperatures, explaining the observed spectral blueshift alongside a quasi-conserved photon rate. Towards the universal temperature, the thermal contribution dominates and a sharp rise in intensity is observed across all frequencies, consistent with thermal emission.
Such behavior of temperature-dependent PL emission (Eq.~(\ref{eq PL_broadband})) is illustrated for various temperatures in Fig.~\ref{Fig1 spectra vs T}. 

In terms of the total rate, the photon flux is obtained by integrating Eq.~(\ref{eq PL_broadband}) over all frequencies:
\begin{equation} \label{eq PL broadband integral}
    \int\limits_{(\nu)}  {{R_{total}}\left( {\nu ,T,{T_p}} \right)d\nu }  = \alpha \left( {1 - QE} \right) \cdot \int\limits_{(v)}  {{R_{BB}}\left( {\nu ,T} \right)d\nu }  + \alpha  \cdot QE \cdot \int\limits_{(\nu)}  {{R_{pump}}\left( {\nu ,{T_p}} \right)d\nu } 
\end{equation}
which is similar to the rate of a 2-level system, shown in Eq.~(\ref{eq Kurtulik: R_th R_pump}). 
As shown in Fig.~\ref{Fig1 rate vs T}, the total emission rate is quasi-conserved at low temperatures and increases with temperature, intersecting at the universal point. While similar behavior has been reported experimentally ~\cite{manor2015conservation, Manor2016, Kurtulik2025}, Eq.~(\ref{eq PL_broadband}) and Eq.~(\ref{eq PL broadband integral}) provide, to our knowledge, the first continuous quantitative model of this transition.

From Eq.~(\ref{eq. Generalized Planck}) and Eq.~(\ref{eq PL_broadband}), the chemical potenital is obtained:
\begin{equation}\label{eq chemical potential}
    \begin{aligned}
    \mu \left( T \right) &\approx kT\ln \left( {\frac{{{R_{PL}}\left( {T,{T_p}} \right)}}{{{\alpha \cdot R_{BB}}\left( T \right)}}} \right)\\ &\approx kT\ln \left[ {1 + QE\left( {\frac{{\int\limits_{(\nu)} {{R_{pump}(T_p)} \cdot d\nu } }}{{\int\limits_{(\nu)}  {{R_{BB}(T)} \cdot d\nu } }} - 1} \right)} \right]
    \end{aligned}
\end{equation}
consistent with prior treatment of $\mu$~\cite{ross1967some,wurfel1982chemical,ries1991chemical,manor2015conservation,Kurtulik2025,yablonovitch1995light}.  However, by implementing Eq.~(\ref{eq PL_broadband}) describing the total thermal and excitation contributions, we show that the chemical potential is expressed as a function of the temperature ($T$), $QE$, bandgap frequency, and pump ($T_p$). By knowing these properties, one can assign any temperature and frequency to PL radiation, analogous to how Planck’s law describes blackbody radiation. 

\begin{figure}[h!]
    \centering
    \begin{subfigure}[]{0.49\textwidth}
        \centering
        \includegraphics[width=\linewidth]{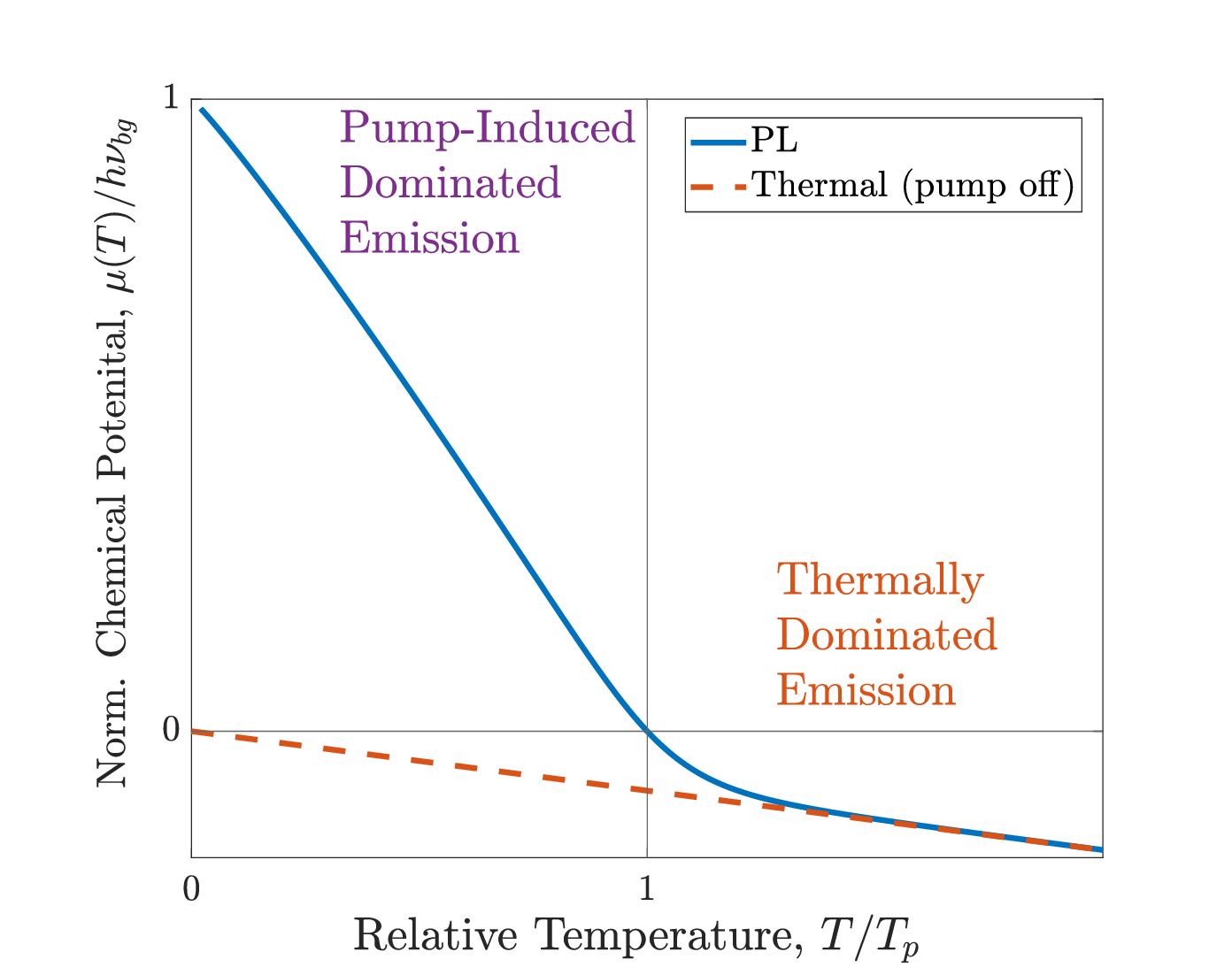}
        \caption{}
        \label{Fig chemical potential}
    \end{subfigure}
    \hfill
    \begin{subfigure}[]{0.49\textwidth}
        \centering
        \includegraphics[width=\linewidth]{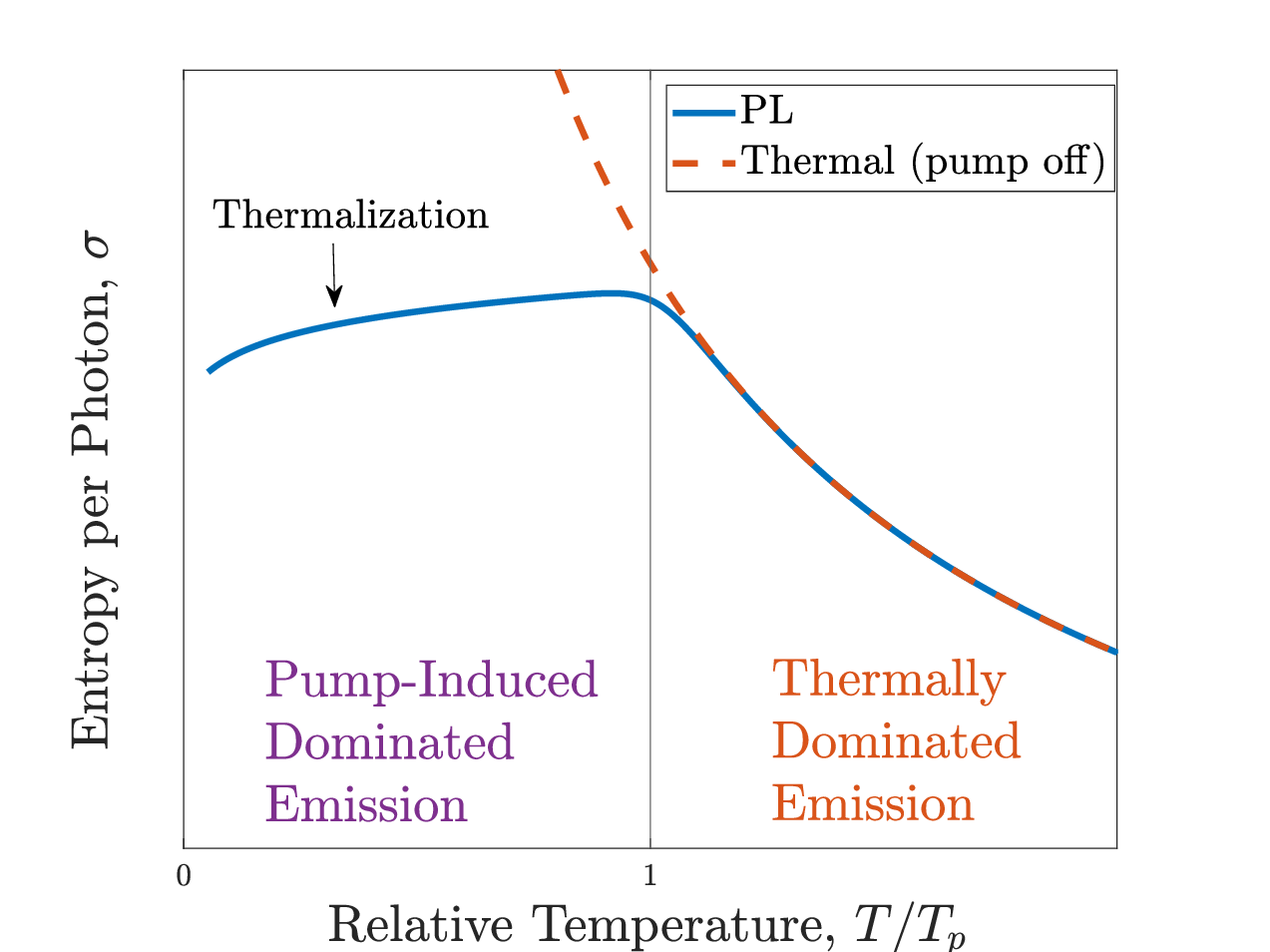}
        \caption{}
        \label{Fig entropy}
    \end{subfigure}
    \caption{Chemical potential and entropy vs temperature, where $T_p=750 [K]$ and $v_{bg} = 2 \times 10^{14}$ [Hz]. (a) Chemical potential of PL emission (blue) and corresponding emission by a similar unpumped material, emitting thermal radiation (dashed). $\mu >0$ signals electronics excitation above thermal, and is reduced with temperature, as zero signals the point of equilibrium (universal point). For $T>T_p$, negative $\mu$ approaches the thermal emission line at high temperatures. (b)  Entropy of PL emission (blue) and unpumped PL material thermal emission (dashed). The entropy increases as thermalization takes place, eventually reaching the universal temperature where the entropy is maximal. Beyond this point, the entropy is reduced, behaving as a thermal emitter.}
\end{figure}

As $T \rightarrow 0$, the chemical potential $\mu$ approaches the bandgap energy $h\nu_{\text{gap}}$ for any material with nonzero quantum efficiency. For $T < T_p$, emission exceeds thermal equilibrium and $\mu > 0$, decreasing with temperature. At the universal point ($T = T_p$), $\mu = 0$ as expected, and for $T > T_p$, $\mu < 0$ as the material emits more energy than it absorbs~\cite{Herrmann2005}. This trend converges toward the chemical potential of an identical, unpumped system, marking the transition to thermally dominated emission, as shown in Fig.~\ref{Fig chemical potential}.

In Eq.~(\ref{eq chemical potential}), the chemical potential accounts for both thermal and pump contributions. In contrast, earlier studies by Yablonovitch on solar cells, where $\mu = qV_{OC}$, neglect the thermal contribution, which is a reasonable approximation given that $T_{\text{cell}} \ll T_{\text{Sun}}$~\cite{ross1967some,Miller2012Yablonovitch}. At low temperatures, however, the two descriptions are effectively equivalent, as the thermal contribution becomes negligible. See Supplement 1 for further discussion.

Deriving the entropy per photon is done directly from the Gibbs free energy~\cite{wurfel1982chemical, manor2015conservation, ross1967some}:
\begin{equation}\label{eq entropy}
    \sigma  = \frac{{h\nu  - \mu }}{T} = {\sigma _{\text{thermal}}} - \frac{{\mu \left( T \right)}}{T}
\end{equation}
Where $\sigma$ is the entropy per photon, equal to the change in entropy for an emitted photon, $\sigma = \Delta S$. While a thermal emitter is characterized by maximal entropy per photon energy, PL exhibits reduced entropy due to the presence of a chemical potential, where the thermalized energy is $h\nu-\mu$~\cite{yablonovitch1995light,manor2015conservation}, as shown in Fig.~\ref{Fig entropy}. 
For $T<T_p$, the entropy increases with temperature due solely to thermalization, indicating an efficient heat engine, transferring the heat into work in the form of blue-shifted photons (optical refrigeration)~\cite{Manor2016}. At the universal point, the chemical potential vanishes and the entropy per photon converges to its thermal value, $\sigma = \sigma _{\text{thermal}}$. 
Above this point, the thermal component dominates and the entropy rapidly decreases, approaching that of the unexcited system (pump = 0), similar to the chemical potential.
It is important to note that this is not identical to pure thermal radiation, where the $QE=0$. 

Next, we analyze how the emission spectrum governs the temporal coherence of PL. The spectral density and temporal coherence function, $\Gamma(\tau)$, are related by the Fourier transform (Wiener-Khinchin theorem) \cite{Davies1990}:
\begin{equation} \label{eq wiener khinchin}
    \Gamma(\tau,\mu) = \int \limits_{(\nu)} R_{PL}(\nu,\mu) \cdot e^{j2\pi\nu\tau}d\nu
\end{equation}

\begin{figure}[h!]
    \centering
    \begin{subfigure}[]{0.49\textwidth}
        \centering
        \includegraphics[width=\linewidth]{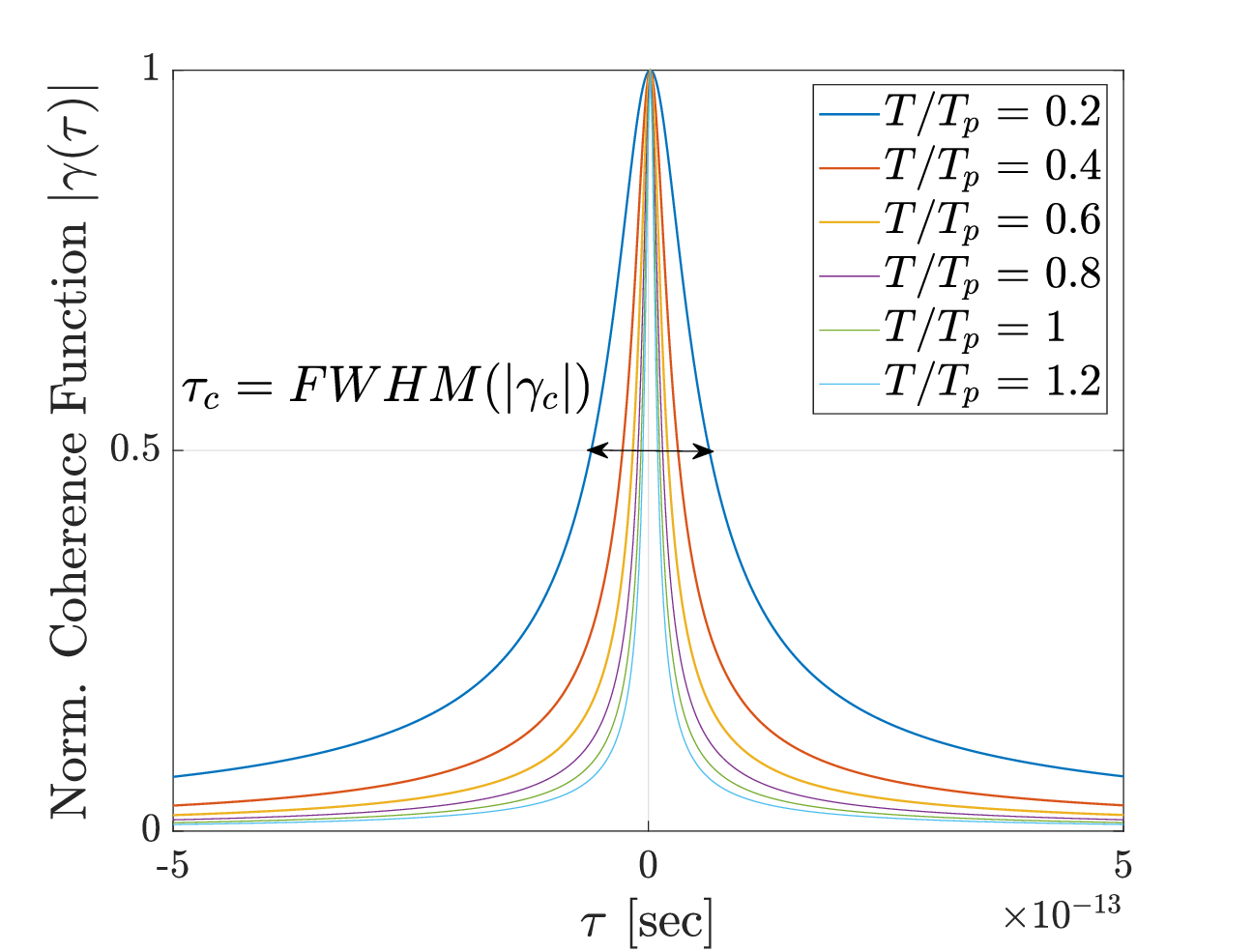}
        \caption{}
        \label{Fig2 Coherence function vs T}
    \end{subfigure}
    \hfill
    \begin{subfigure}[]{0.49\textwidth}
        \centering
        \includegraphics[width=\linewidth]{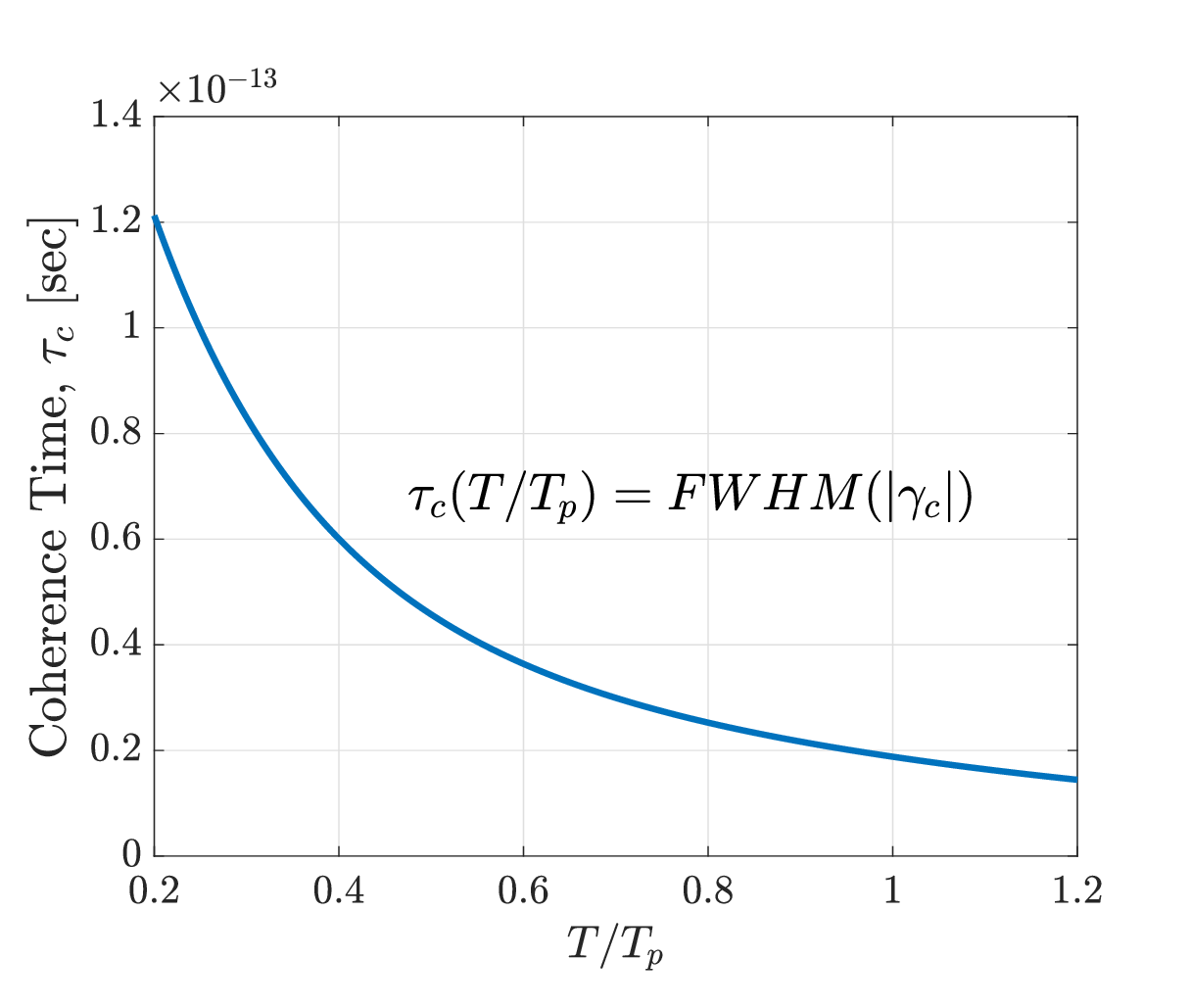}
        \caption{}
        \label{Fig2 CoherenceTime vs T}
    \end{subfigure}
    \caption{Results for a PL body at different temperatures, where $T_p = 1000$ [k], $\nu_{bg} = 10^{14}$ [Hz], corresponding with 1.5 microns. (a) Normalized temporal coherence function of PL material for different relative temperatures, where $T/T_p=1$ is the universal temperature and the coherence time equals the full-width-half-maximum. (b) Coherence time vs. relative temperature. Inversely proportional to the temperature. At low temperatures, PL emits with high intensity and narrow bandwidth, therefore having a longer coherence time and length.}
\end{figure}

The normalized temporal coherence function is, therefore:
\begin{equation}\label{eq norm coherence gamma}
    |\gamma(\tau,\mu)| = \left | \frac{\Gamma(\tau,\mu)}{\Gamma(0,\mu)} \right |
\end{equation}
which is directly associated with the first-order coherence function, $g^{(1)}(\tau)$, and the visibility function, $V$, \cite{Fox2006}. Conventionally, the coherence-time equals the full-width-half-maximum (FWHM) of $|\gamma(\tau)|$. Fig.~\ref{Fig2 Coherence function vs T} depicts the evolution of the normalized temporal coherence function for material having a bandgap at $2 \times 10^{14}$ Hz $(1.5 \mu m)$, while Fig.~\ref{Fig2 CoherenceTime vs T} depicts the evolution of the coherence time with temperature. As expected, the coherence time is inversely proportional to temperature and depends on the bandgap. As shown in Eq.~(\ref{eq PL_broadband}), the PL spectral width depends only on the material’s temperature and bandgap; thus, the coherence time is independent of the pump and $QE$. Any $QE$ value material would have a similar normalized coherence function, $\gamma(\tau)$, as a gray-body having the same bandgap. This understanding provides a foundation for temperature-controlled light sources, in which the coherence time and length are determined by the temperature and bandgap, while the emission intensity depends on the pump rate and $QE$.

We now discuss the evolution of photon statistics. When limited to a single mode of radiation with energy $h\nu$, different emission mechanisms may exhibit different probabilities of emitting photons within a given time interval, which depend on the quantum state of the system \cite{Fox2006,Loudon2000}. For example, a coherent state obeys Poissonian statistics while a thermal state obeys Bose-Einstein statistics (super-Poissonian). Here, we develop the evolution of PL photon statistics with temperature. Following Pearsall~\cite{Pearsall2021}, for a PL emitter characterized by a defined temperature and chemical potential, the mean photon number is given in the generalized Planck’s law:
\begin{equation}\label{eq BE photon statistics}
    \langle n(h\nu,\mu) \rangle = \frac{1}{e^{\frac{h\nu-\mu}{kT}}-1}    
\end{equation}
Using the power of statistical mechanics, the variance for any $\mu$ is calculated as follows :
\begin{equation}
\begin{aligned}\label{eq Var_n}
    \text{Var}(n(h\nu,\mu)) &= kT \frac{\partial}{\partial \mu}\langle n\rangle
    \\
    &= \langle n\rangle^2 + \langle n\rangle
\end{aligned}
\end{equation}
obeying Bose-Einstein statistics, similar to thermal light. Thus, the second-order coherence function, $g^{(2)}$, is related to $\gamma(\tau,\mu)$  from Eq.~(\ref{eq norm coherence gamma})
 via the Siegert relation~\cite{Siegert1943,Loudon2000}:
\begin{equation}
    g^{(2)}(\tau,\mu) = 1 + |\gamma(\tau,\mu)|^2
\end{equation}
and reduces to $g^{(2)}(\mu,0)=2$,  reflecting the tendency of photons to bunch together. 
Doronin \textit{et al.} have reported similar results through simulations, demonstrating that the light produced by amplified spontaneous emission (ASE), even below threshold, which is consistent with PL, exhibits photon statistics characteristic of thermal light, with $g^{(2)}(0) = 2$~\cite{doronin2019second}.  
Additionally, experimental work by Valero \textit{et al.} shows that ASE with different coherence times also exhibits thermal photon statistics. Although not tested directly on PL, in the pump-induced regime where $T \ll T_p$, population inversion can lead to ASE in PL systems~\cite{valero2021high}.

\begin{figure}[h!]
\centering
\includegraphics[width=\linewidth]{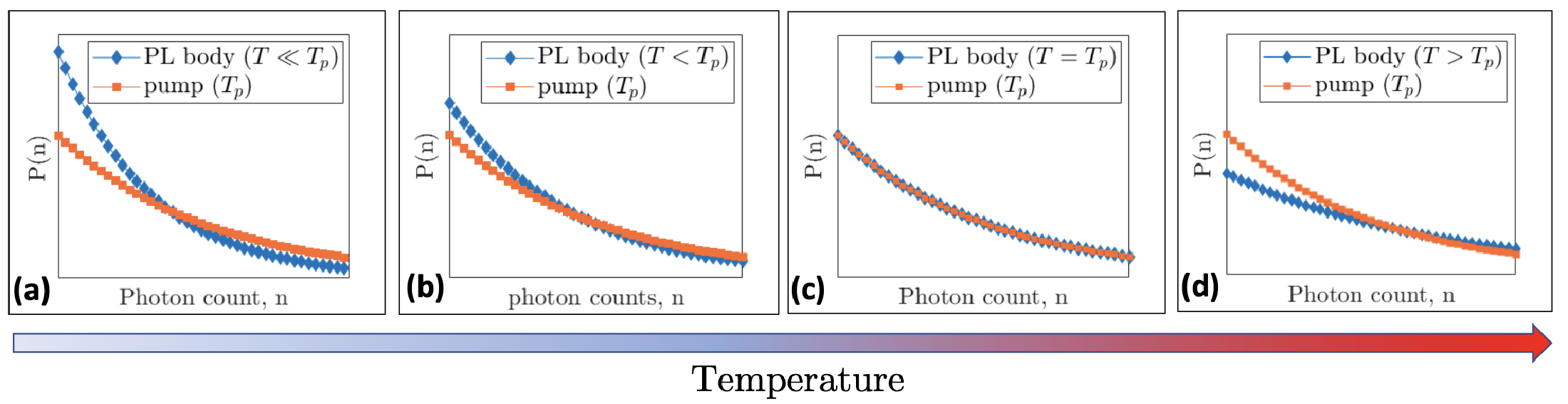}
\caption{Photon statistics comparison between PL body at different temperatures and a constant thermal pump. BE statistics is demonstrated at four different regimes: $T \ll Tp$, $T < T_p$, $T = T_p$ (universal temperature), and $T > Tp$. (c) is the equilibrium point since the material and pump share an identical distribution.}
\label{Fig4 photon statistics}
\end{figure}

When the pump is thermal and $T<T_p$, a temperature difference exists between the pump and the PL material, causing the two emission sources to follow different thermal statistical distributions (Fig.~\ref{Fig4 photon statistics}). However, upon reaching the universal point, they share identical thermal statistics, where all moments become equal Fig.~\ref{Fig4 photon statistics}. Therefore, due to the thermalization and equilibrium between the electrons and the phonons in the material, the PL, from a statistical perspective, behaves like a thermal emitter at the body's temperature.

When confined to a single mode of radiation, PL can yield high-intensity emission with thermal statistics. As shown in Eq.~(\ref{eq BE photon statistics}) and Eq.~(\ref{eq Var_n}), the intensity is determined by the temperature and chemical potential, even under strong non-thermal excitation such as laser pumping, due to carrier thermalization.


Throughout this analysis, we have discussed the case of a thermal pump at $T_p$. 
Addressing the conditions for equilibrium between a non-thermal pump, such as a laser, and a PL body presents a different scenario. The solution to this problem may be approached in two ways. From the statistical perspective of a single radiation mode, even if the intensities are matched, the system cannot share the same variance due to the inherent difference in photon statistics. Unlike a coherent state, a thermal state exhibits enhanced fluctuations (i.e., greater variance). 
Alternatively, a thermodynamic perspective can also be adopted. For thermodynamic equilibrium, we require equal entropy generation by the PL emitter and the pump. Würfel shows that the entropy of a laser tends towards zero as $\mu_{laser} \rightarrow h\nu$ \cite{wurfel1982chemical}, therefore:
\begin{equation}\label{eq laser-PL equilibirium}
    \sigma = \frac{h\nu-\mu}{T} = \frac{h\nu-\mu_{laser}}{T_p} = 0
\end{equation}
For any material with $QE<1$, maintaining energy balance under optical pumping requires a temperature increase. This results in entropy generation and a corresponding reduction in the chemical potential, in contradiction with the condition set by  Eq.~(\ref{eq laser-PL equilibirium}).

\section{Conclusion}
This study presents what is believed to be the first analytical model that quantitatively relates the chemical potential of PL radiation to temperature, quantum efficiency $(QE)$, bandgap energy, and external excitation. Unlike previous studies that focus on specific materials, this work aims to provide a generalized description of temperature-dependent PL.
We show that at low temperatures $(T < T_p)$, the emission rate remains quasi-conserved, as the chemical potential decreases and the entropy increases, while at elevated temperatures $(T > T_p)$, all three quantities undergo a rapid transition to thermal behavior. In contrast, the coherence time and photon statistics evolve smoothly across the entire temperature range.
At the universal temperature $(T = T_p)$, thermodynamic equilibrium is established such that the PL emission rate equals the absorbed pump rate, and the photon statistics matches that of the thermal pump. Finally, we discuss the case of a non-thermal pump and argue that no equilibrium can be reached.

Beyond its fundamental contribution, this model provides a framework for designing tunable light sources, where the coherence time and intensity are controlled by the temperature and pump excitation, respectively. It also supports the design of efficient sources that exhibit thermal photon statistics at high intensities.

\textbf{Funding.} Israel Science Foundation (1230/21).

\textbf{Disclosures.} The authors declare no conflicts of interest.

\textbf{Data availability.} Data underlying the results presented in this paper are not publicly available at this time but may be obtained from the authors upon reasonable request.

\textbf{Supplemental document.} See Supplement 1 for supporting content.


\bibliographystyle{unsrt}  
\bibliography{references}

\end{document}